	\documentstyle{article}%
\begin{document}

\title{Nonperiodic Orbit Sums in Weyl's Expansion for Billiards}
\author{{\it Wei-Mou Zheng} \\
{\it Institute of Theoretical Physics, Academia Sinica, Beijing 100080, China%
}}
\date{}
\maketitle

\begin{abstract}
Weyl's expansion for the asymptotic mode density of billiards consists of the 
area, length, curvature and corner terms. The area term has been associated 
with the so-called zero-length orbits. Here closed nonperiodic paths 
corresponding to the length and corner terms are constructed. 
\end{abstract}

\leftline{PACS numbers: 05.45., 03.65.Sq}%

\section{ Introduction}

The asymptotic mode density appears in various branches in physics. In a
very early analysis of the density of eigenmodes for a cavity with reflecting 
walls, Weyl proved that the leading term is proportional solely to the volume
of the domain, and independent of the shape \cite{weyl}. Since then terms
which account for the surface, shape and connectivity of the domain
were found to refine asymptotic expansions \cite{bb,b&h,kac}. For a billiard
inside a simply connected domain of the Euclidean plane, the two-dimensional
version of Weyl's expansion reads
\begin{equation}
\rho (E)\sim \frac {\cal A}{4\pi} -\frac {\cal L}{8\pi \sqrt{E}}+\left[
\frac 1{12\pi}\oint c(s)\,ds+\frac 1 {24}\sum_i\left(\frac \pi{\alpha_i}
-\frac {\alpha_i}\pi\right)\right]\delta (E),
\label{weyl}
\end{equation}
where Dirichlet boundary conditions are used, units are set at $2m=\hbar
=1$, the boundary is characterized by the arc length $s$, and ${\cal A}$,
${\cal L}$, $c(s)$, $\alpha_i$ are the area, the total length, curvature and
inner angle of corner, respectively.

Most studies of eigenvalue densities are based on the Green's function
technique. Mathematical asymptotic theory involves the Tauberian theorems
from the theory of Laplace transforms. Balian and Bloch worked with energy
Green's functions. Without requiring Tauberian theorems, they obtained a
multiple reflection expansion for Green functions. The use of curvilinear
coordinates was also included in their discussions.

The area term has the simple meaning that the probability for a system to be
in a particular subset of phase space is proportional to the volume of the
subset. This term has been associated with the so-called zero-length orbits
in the semiclassical theory \cite{b&m}. 

The periodic orbit theory of Gutzwiller, known as the trace formulas,  relates
the fine details of oscillating density of states with classical periodic orbits
\cite{gutz, brack}. A trace formula has been derived for integrable systems by
Berry and Tabor \cite{b&t}. Other extensions of the Gutzwiller theory have
been developed \cite{stru,ozo}. However, not much attention has been drawn to
the role of nonperiodic orbits. Here we shall relate some nonperiodic orbits
to Weyl's expansions. In Sec.~II a family of closed nonperiodic orbits of 
single reflection is given to account for the length term. In Sec.~III orbits
contributing to the corner term are examined. Finally, some remarks are made
in the last section.

\section{ The length term}

Using the stationary method, Gutzwiller derived from Feynman's path integral
the approximate propagator from ${\bf r}$ to ${\bf r}'$ in the time difference 
$t$ \cite{gutz}
\begin{equation}
K_{sc}({\bf r},{\bf r}';t) = \sum_{\rm cl.tr.}(2\pi i)^{-N/2} \sqrt{|\det
C|} \exp\left[i R({\bf r},{\bf r}';t) -iM\pi /2\right],
\label{k}
\end{equation}
where $R$ is Hamilton's principal function, $M$ the phase index obtained by
counting the number of conjugate points along each classical path over which
the summation is taken, $N$ the dimensionality, and $C$ the matrix of the
negative second variations of $R$
\begin{equation}
C_{ij}({\bf r},{\bf r}';t) =-\frac {\partial^2R}{\partial r_i\partial r'_j}.
\end{equation}
By converting time to
energy, the propagator is transformed to Green's function as
\begin{equation}
G({\bf r},{\bf r}';E) =i^{-1}\int_0^\infty dt\, K({\bf r},{\bf r}';t) 
\exp (iEt) .
\label{k-g}
\end{equation}
By using the propagator (\ref{k}), after integrating over time again by the
method of stationary phase, Gutzwiller's semiclassical approximation to
Green's function is given by
\begin{equation}
G_{sc}({\bf r},{\bf r}';E) = \frac{2\pi}{(2\pi i)^{(N+1)/2}}\sum_{\rm cl.tr.}
\sqrt{|D|}\exp \left[iS({\bf r},{\bf r}';E) -i\mu\pi /2\right],
\label{g}
\end{equation}
where $S$ is the action integral, $\mu$ is again the phase index, i.e. the 
number of conjugate points, but obtained by varying the trajectory at constant
energy $E$ instead of time $t$, and
\begin{equation}
D= (-1)^N \frac{\partial^2S}{\partial E^2}\det C =(-1)^N\frac 1{\dot q\dot q'}
\det\left(\frac {\partial{\bf p}'_\perp}{\partial{\bf r}_\perp }\right),
\label{d}
\end{equation}
where in the last equation ${\bf r}_\perp$ is in the subspace transverse to
the trajectory in the local coordinate system, and ${\bf p}_\perp$ is its 
conjugate momentum. Once Greeen's function is known, the density of states 
can be calculated by
\begin{equation}
\rho(E)=-\frac 1\pi{\rm Im}\int d{\bf r}\,G({\bf r},{\bf r};E).
\label{rho}
\end{equation}

When the semiclassical approximation is made, the density of states is
expressed in terms of classical closed orbits. The zero-length orbits
contribute to the average density of states. When the
method of stationary phase is again used for the integration over ${\bf r}$,
only the periodic orbits still remain in the summation.

For a billiard, there is a family of closed orbits involving single
reflection. Consider the simplest case when the boundary consists of the
$x$-axis, and the inside region is the upper plane. A closed orbit of the
family is an orbit going from $(x,y)$ (with $y$ positive) to $(x,0)$ and
then returning back to $(x,y)$. The calculation of the Jacobian $\partial
r_\perp/\partial p'_\perp$ in the expression (\ref{d}) for D is given in
the appendix. From Eqs.~(\ref{d}), and (\ref{j0}), noticing that
$$E\equiv k^2,\quad L\equiv 2y,\quad \dot q=2k=L/t, \quad \partial^2S/
\partial E^2=-L/(4k^3),$$
we find
\begin{equation}
D= 1/(8ky), \quad |\det C|= 1/(4t^2).
\label{d&c}
\end{equation}
It is easy to verify that Hamilton's principal function $R$ and the action 
integral $S$ are
\begin{equation}
R=y^2/t, \qquad S=2ky.
\end{equation}
From the approximation (\ref{k}) for the propagator, we have
\begin{equation}
K_{\rm co} = -\frac 1{4i\pi t}\exp \left(i\frac {y^2}t\right),
\label{kco}
\end{equation}
where subscript `co' stands for `closed orbits'. By using the formula for 
Hankel's function
\begin{equation}
H_\nu ^{(1)}(xz)= \frac{i^{-\nu -1}}{\pi} z^\nu \int_0^\infty \exp\left[
\frac 1 2 ix\left(t+\frac {z^2} t \right)\right] t^{-\nu -1}\,dt,
\label{H}
\end{equation}
from Eq.~(\ref{k-g}) Green's function is obtained as
\begin{equation}
G_{\rm co}({\bf r},{\bf r};E)= -\frac 1 {4i}H_0^{(1)}(2ky).
\label{gco}
\end{equation}
By means of the relation (\ref{rho}), its contribution to the density of states 
may be estimated by using
\begin{equation}
\int_0^\infty dz\, z^\mu H_\nu^{(1)}(az)=\frac 1\pi 2^\mu i^{\mu -\nu}a^{-\mu 
-1} \Gamma \left(\frac{1+\mu +\nu}2\right)\Gamma \left(\frac{1+\mu 
-\nu}2\right)
\label{H_int}
\end{equation}
to be 
\begin{equation}
-\int dx\int_0^\infty dy\,\frac 1 {4i}H_0^{(1)}(2ky)=-\frac {\cal L}{8\pi 
\sqrt{E}},
\end{equation}
which gives exactly the length term in Weyl's expansion (\ref{weyl}).

If we estimate Green's function directly from Eq.~(\ref{g}), the result becomes
\begin{equation}
G'_{\rm co} = -\frac 1{4i\sqrt{\pi ky}}\exp (2iky),
\end{equation}
which is consistent with the above $G_{\rm co}$ in the asymptotic approximation. However, its contribution to the density of states is ${\cal L}/(4\sqrt{2E}\pi )$, which is different from the expected result by a factor of $\sqrt{2}$. This situation is similar to that of zero-length orbits. Since integration over small $y$ contributes significantly, we have to use the uniform approximation 
with Hankel's function.

When a bounce takes place at a position $s$ of the boundary with curvature 
$c(s)$, the closed orbit along the normal to the boundary still exists. Set 
the coordinate system with the origin at the bounce point and the $y$-axis 
along the normal directing towards the inside of the billiard. By using 
Eqs.~(\ref{m}) and (\ref{j}), the counterparts of quantities (\ref{d&c}) are 
found to be
\begin{equation}
D= 1/[8ky(1-cy)], \quad |\det C|= 1/[4t^2(1-cy)],
\label{circ:d&c}
\end{equation} 
which gives the same length term.

\section{The corner term}
In the previous section we have considered closed orbits of single reflection. 
For a corner with an acute angle there is a family of closed orbits with 
double reflection. We shall first examine an acute corner, and then extend the 
analysis to an obtuse corner. 

\subsection{Acute corners}
For a corner with an acute angle $\alpha$ we may construct a closed orbit of 
double reflection as follows. Let us denote by $O$ the vertex of the corner, 
and by $OA$ and $OB$ its two sides. Suppose that the mirror image of $OA$ with 
respect to $OB$ is $OA'$, and the image of $OB$ with respect to $OA'$ is $OB'$. 
In this way we get the first and second images of the original corner, and a 
given point $Q$ is then mapped to $Q_2$ in the corner $A'OB'$, as shown in 
Fig.~1. The straight line $QQ_2$, after mapping its segments back into the 
original corner, gives the closed orbit looked for. 

Denote by $r=|OQ|$ the length of $OQ$. The length of the closed orbit is then
\begin{equation}
|{\rm QQ}_2|=2r\sin \alpha .
\end{equation}
From Eqs.~(\ref{j0}) and (\ref{k}), we have
\begin{equation}
K_{\rm co2}= \frac 1{4i\pi t}\exp\left(i\frac{(r\sin\alpha )^2}t\right),
\label{kco2}
\end{equation}
which, from Eqs.~(\ref{k-g}) and (\ref{rho}), leads to
\begin{equation}
\rho_{\rm co}=\frac \alpha {8\pi\sin^2\alpha}\delta (E) ,
\label{cor}
\end{equation}
where we have used formula (\ref{H_int}) and have written $E$ as $E+i\epsilon$ 
to give
$$\frac 1{E+i\epsilon} ={\cal P}\frac 1E +\frac\pi i \delta (E).$$
The corner term in Weyl's expansion for the rectangular corner is estimated to 
be 1/16, in agreement with the value given by expression (\ref{cor}). 

Expression (\ref{cor}) is not the same as the corner term in Weyl's expansion 
(\ref{weyl}). A corner gives a correction to the length term due to the 
restriction on the domain of integration. Set the vertex O at the origin, and 
side OA along the $x$-axis. The change in the domain of integration
$$\int dx\int_0^\infty dy \quad\to\quad \int dx\int_0^{\gamma x}dy, {\rm 
\quad with\quad} \gamma = \tan\alpha $$
gives the correction to the density of states
\begin{equation}
\delta\rho = -\frac 2\pi{\rm Im}\int_0^\infty dy \int_0^{y/\gamma} dx\,
\frac 1 {4i}H_0^{(1)}(2ky) = \frac 1{4\pi\gamma}\delta (E),
\end{equation}
where the factor 2 in front accounts for the two sides, and use of Green's 
function (\ref{gco}) for single reflection orbits has been made. Thus, by 
combining the above two corrections, the total is
\begin{equation}
\rho_{\rm co2}= \frac 1{8\pi}\left(\frac \alpha {\sin^2\alpha}+2\cot\alpha
\right)\delta (E).
\label{acute}
\end{equation}
This is not of the same form as that in Weyl's expansion. For $\alpha =(1/2 
-\delta )\pi$ a little smaller than the rectangle, from Eq.~(\ref{acute}) the 
lowest order is $1/16+\delta /8$, while that from Weyl's expansion is $1/16+ 
5\delta /24$. For very small $\alpha$, compared with $\pi /(24\alpha )$ from 
expression (\ref{weyl}), the semiclassical value is $3/(8\pi\alpha )$. The 
ratio is $9/\pi^2 \approx 0.912$.

In the above discussion we have considered only orbits of double 
reflection which hit the side $OB$ first. An extra factor 2 should appear 
for those orbits to count the two different ways of 
selecting orbits according to which side is hit first.
 
\subsection{Obtuse corners}
For an obtuse corner the above closed orbits of double reflection do not exist. 
A natural way of continuation has to be found. For this purpose we use the 
folding property of the propagator
\begin{equation}
K({\bf r},{\bf r}';t-t')=\int d{\bf r}''\,K({\bf r},{\bf r}'';t-t'') 
K({\bf r}'',{\bf r}';t''-t'),
\end{equation}
to include two-piece closed orbits. We first examine the case of an acute 
corner. Let us consider a two-segment broken line from $Q$ to $Q_2$ via some 
mediate point $Q'$ shown in Fig.~1, and set $t-t''=t''-t'=\tau$. Suppose that 
the polar coordinates of these three points are $(r,\theta_1)$, $(r_2,\theta_2)
=(r,2\alpha +\theta_1)$ and $(r_0,\theta_0)$, respectively. In similarity to 
Eq.~(\ref{kco}), we may derive the propagator for each segment. By means of 
the folding property, we find for the broken path
\begin{equation}
K(Q,Q_2|Q') =\frac 1{4i\pi\tau )^2}\int d\theta_0 \exp\left\{ \frac i{4\tau}
\left[2r^2+2r_0^2-2rr_0\cos(\theta_0-\theta_1)-2rr_0\cos(\theta_0-
\theta_2)\right]\right\},
\label{kqq"}
\end{equation}
where the integration domain for $\theta_0$ is determined by the constraints
$0\leq\theta_0\leq 3\alpha$, $0\leq\theta_1\leq\alpha$, $|\theta_0-\theta_1|
\leq\pi$ and $|\theta_0-\theta_2|\leq\pi$. At the limit of the rectangular 
corner, propagator (\ref{kqq"}), after integrating over $\theta_1$, reduces to
\begin{equation}
\int_0^\alpha d\theta_1\,K(Q,Q_2|Q')=\frac \pi 2 K(Q,Q_2|Q')=
\frac \pi 2\left\{\frac 12\frac 1{4i\pi t}\exp\left(i\frac {r^2} t\right)
\right\}.
\label{fold}
\end{equation}
Compared with the expected form (\ref{kco2}), $K(Q,Q_2|Q')$ equals the half of 
$K_{\rm co2}$. It can be verified that if we use the method of stationary phase 
approximation in the Cartesian coordinate system for the integration involving 
$Q'$ the propagator (\ref{kqq"}) would revert to (\ref{kco2}) exactly. However, 
if we count the two ways of selecting orbits, the result here turns out to be 
better.

Generally, we may approximate the propagator between a reference point and 
some mediate one, say $Q$ and $Q'$ in Fig.~1, by the semiclassical propagator 
involving classical paths with 0 to 2 bounces. Besides images $Q_1$ and $Q_2$, 
point $Q$ has two more images $Q_{-1}$ and $Q_{-2}$ obtained clockwisely, as 
shown in Fig.~1. The paths which contribute to the semiclassical propagator 
from $Q'$ to $Q$ are straight segments $Q'Q$, $Q'Q_1$, $Q'Q_2$, $Q'Q_{-1}$ and 
$Q'Q_{-2}$. The last two correspond to paths hitting side $OA$ first. 
Similarly, by considering images of $Q'$, paths contributing to the propagator 
from $Q$ to $Q'$ can be found. By means of the folding property, combination 
of these two propagators gives an approximate propagator from $Q$ to $Q$ 
itself, which includes $K(Q,Q_2|Q')$ as a part. 

Calculation in the Cartesian coordinate system for the rectangular corner is 
rather easy. In this case both $Q_2$ and $Q_{-2}$ are in the third quadrant, 
while $Q_1$ and $Q_{-1}$ are, respectively, in the second and fourth quadrants. 
Each path arises as a length square in the exponent of the expression for 
propagator. Denote by $(x,y)$ and $(x_0,y_0)$ the coordinates of $Q$ and $Q'$, 
respectively. There is a correspondence between closed paths and length square 
sums $[(x\pm x_0)^2+(y\pm y_0)^2]+[(x\pm x_0)^2+(y\pm y_0)^2]$. Here the first 
square bracket corresponds to the path from $Q$ to $Q'$, and the second to the 
return path. A plus sign between $x$ and $x_0$ indicates a bounce on side $OB$, 
while that between $y$ and $y_0$ indicates a bounce on side $OA$. Thus, each 
path has its `four signs' signature, which is the four signs appearing in the 
square sum. For example, the closed path without any bounces may be marked as 
$----$, which contributes to the area term. It can be verified that each of 
the paths $-+-+$ and $+-+-$ results in 
$$\frac 12\frac {\cal L}{8\pi \sqrt{E}}-\frac 1 {16\pi^2}\delta (E)$$
for the level density, while each of the paths $+---$, $-+--$, $--+-$ and 
$---+$ gives
$$- \frac 12\frac {\cal L}{8\pi \sqrt{E}}+\frac 1 {32\pi}\delta (E).$$
Closed paths with a total of 2 bounces, besides paths $-+-+$ and $+-+-$, are 
$+--+$, $-++-$, $++--$ and $--++$, each of which contributes $\delta (E)/64$. 
Paths with 3 bounces are $+++-$, $++-+$, $+-++$ and $-+++$, each of which 
contributes $-\delta (E)/32\pi$. The only path with 4 bounces is $++++$, whose 
contribution is $\delta (E)/(16\pi^2)$. The total contribution of these 16
terms recover exactly the area and length terms, and gives the corner term as
$$\left(\frac 1 {16}-\frac 1{16\pi^2}\right)\delta (E).$$
We see that the paths with a single bounce on both sides gives the main 
contribution to the corner term. 

Thus, for an obtuse corner, although closed classical orbits of double 
reflection generally do not exist, we may still calculate the corner term 
from the two-piece paths of just 2 single reflections on both sides, which  
are made either by only one piece, or by each of the two pieces. Due to the 
cancellation among terms for the folding propagator the approximation  
keeps the main contribution. However, now the integrals involved cannot be 
estimated analytically, and numerical methods have to be used.
 
\section{Discussions}

In the above we have examined the role played by closed classical orbits with 
single and double reflection on the boundary for billiards. Here we make some 
concluding remarks.

\begin{enumerate}
\item So far, we have considered only the length term and the corner term, 
leaving the curvature term untouched. The same sign for both Dirichlet and 
Neumann conditions gives us a hint that orbits of double reflection is 
dominant. In principle, we can consider the contribution from two-piece closed 
pathes, as we did for obtuse corners. Unfortunately, even the simple case of 
a circle does not admit simple expressions. One way to get round the intricate 
situation of a circle is to consider its inscribed polygons \cite{kac}.

\item A way to derive the propagator for a corner is to use curvilinear 
coordinates. A corner can be `flattened' by introducing the tranformation 
from $(x,y)\to (u,v)$ defined by 
\begin{equation}
x=r\cos(\bar\gamma\varphi ), \quad y=r\sin(\bar\gamma\varphi ), \quad{\rm with}
\quad r^2=u^2+\gamma^2v^2,\quad \tan\varphi =\frac{\gamma v} u ,\quad \gamma 
=\frac \pi \alpha ,\quad \bar\gamma =\frac \alpha \pi ,
\end{equation}
where $\alpha$ is the inner angle of the corner as before. The Jacobian of 
this transformation is equal to one, which is made to keep the area and length 
terms unchanged. The transformed Laplacian can be derived as
\begin{equation}
\Delta =\partial_u^2\partial_v^2 (\gamma^2-1)\left(\partial_u^2-\frac{u^2}
{r^2}\partial_u^2 -\frac{v^2}{r^2}\partial_v^2 -\frac{2uv}{r^2}\partial_u
\partial_v -\frac{u}{r^2}\partial_u -\frac{v}{r^2}\partial_v \right).
\end{equation}
We now regard the product with the factor $(\gamma^2-1)$ as the perturbation 
to  $\partial_u^2\partial_v^2\equiv \Delta_0$. The factor $(\gamma^2-1)$ is 
indeed tempting if one notes that the corner term is
\begin{equation}
\frac {\pi^2-\alpha^2}{24\pi\alpha} =\frac 1{24}(\gamma -\bar\gamma) 
=\frac{\gamma^2-1} {24\gamma}.
\end{equation}
By means of the perturbation expansion for the propagator the corner term can 
be obtained.

\item Only Dirichlet boundary conditions have been considered in the above. 
The extension to Neumann conditions is rather straightforward. Since there 
seems to be no general formula for the corner term at Neumann conditions in 
the literature \cite{brack}, a semiclassical estimation can now be made using 
our approach.

\item The closed paths considered above have a zero limiting length. There 
are other closed orbits of a non-zero limiting length. For example, there is 
a continuous family of closed orbits from the diameter orbit to the 
equilateral triangle orbit in a circular disk. Any member of the family is an 
isosceles triangle with one vertex inside the circle. The role played by 
such orbits is worth examining.  

\item We may extend our analysis to include connectivity and higher 
dimensionality.
\end{enumerate}

Some problems are under study.

\appendix
\section*{Appendix: Jacobian $\partial r_\perp/\partial p'_\perp$}

For a billiard inside a simply connected domain of the Euclidean plane, we
may derive the Poincar\'e map from bounce to bounce in Birkhoff coordinates 
$(s,v)$, where $v$ is the component of the velocity in the tangent direction 
to the boundary right after reflection, and $s$ the arc length along the 
boundary. Since the absolute value of velocity is conserved for a billiard, we
may normalize the velocity as a unit vector and then let $v\in [0,1]$.
The linearized Poincar\'e map from $(s_1,v_1)$ to $(s_2,v_2)$ can be
expressed as \cite{berry, sieb, dul}
\begin{eqnarray}
M(12) &=& \left(
	\begin{array}{cc}
    (l_{12}c_1- v_{1\perp})/v_{2\perp} & -l_{12}/v_{1\perp}v_{2\perp}\\
    c_1v_{2\perp}+ c_2v_{1\perp} -l_{12}c_1c_2&
    (l_{12}c_2- v_{2\perp})/v_{1\perp}
    \end{array} \right)\\
&=& \left(  \begin{array}{cc} 1/v_{2\perp} &0\\ 0 &v_{2\perp}
    \end{array} \right)
    \left(  \begin{array}{cc} 1& 0\\ -c_2/v_{2\perp} &1
    \end{array} \right)
    \left(  \begin{array}{cc} -1& -l_{12}\\ 0& -1
    \end{array} \right)
    \left(  \begin{array}{cc} 1& 0\\ -c_1/v_{1\perp} &1
    \end{array} \right)
    \left(  \begin{array}{cc} v_{1\perp}& 0\\ 0 &1/v_{1\perp}
    \end{array} \right) ,
\end{eqnarray}
where $l_{12}$ is the length of the chord joining $s_1$ and $s_2$, 
$v_\perp$ is the normal component of the velocity and $c$ the curvature
of the billiard boundary. It is often useful to know the Jacobian matrix
$\partial (s,v)/\partial (\xi ,\kappa )$, where $\xi$ and $\kappa$ are
perturbations in displacement and velocity at a given point O on the straight
line joining $s_1$ and $s_2$ along the direction perpendicular to the path.
Without loss of generality we may choose the coordinate system with the origin
at O, and the $y$-axis along the path from $s_1$ to $s_2$, as shown in
Fig.~2. (In the figure $s_1$ and $s_2$ are marked as $s$ and $s'$,
respectively.) It is obvious that at $s_1$
\begin{equation}
v_x=0,\qquad v_y=1.
\label{vx0vy1}
\end{equation}
Denote by a dot the derivative with respect to the arc length $s$. The tangent
and normal unit vectors are ${\bf t}_1=(\dot x_1, \dot y_1)$ and ${\bf n}=
(-\dot y_1, \dot x_1)$, respectively. This implies that
$$v_1=\dot x_1 {\quad \rm and \quad} v_{1\perp}=\dot y_1 .$$
Assume that a perturbation $(\delta s, \delta v)$ from
$s_1$ to $\tilde s=s_1+\delta s$ results in the perturbation at O,
$(\xi ,\kappa )$. Up to the lowest order, we have
\begin{equation}
\tilde x\equiv x(\tilde s) =x_1+\dot x_1\delta s= v_{1\perp}\delta s {\quad\rm and\quad}
\tilde y\equiv y(\tilde s)=y_1+\dot y_1\delta s\approx y_1.
\label{x'y'}
\end{equation}
It can be seen that $\delta v_{1x}\equiv {\tilde v}_x-v_{1x}=\kappa$. From Eq.~(
\ref{vx0vy1}), the relation $v_x(\delta v_x)+v_y(\delta v_y)=0$ implies that
$\delta v_{1y}=0$. That is, at $\tilde s$ we have ${\tilde v}_x=\kappa$, and ${\tilde v}_y=
v_{1y}=1$. Using ${\tilde v}_x/{\tilde v}_y=(\tilde x-\xi )/\tilde y$, we find
\begin{equation}
v_{1\perp}\delta s_1 -\xi =y_1 \kappa .
\label{ds}
\end{equation}
Similarly, the relation $v=v_x\dot x +v_y\dot y$ leads to
\begin{equation}
\delta v_1= v_{1\perp}\kappa +\ddot y\delta s= v_{1\perp}\kappa 
+c_1 v_{1\perp}\delta s,
\label{dv}
\end{equation}
where we have used the curvature formula $\ddot y=c\dot x$. Equations
(\ref{ds}) and (\ref{dv}) can be written as
\begin{equation}
\left(\begin{array}{c} \delta s_1 \\ \delta v_1 \end{array} \right)
=\left( \begin{array}{cc}
 1/v_{1\perp}& y_1/v_{1\perp}\\ c_1 &v_{1\perp}+c_1y_1 \end{array} \right)
 \left( \begin{array}{c} \xi \\ \kappa \end{array} \right).
\end{equation}
The above transform matrix may be written as
\begin{equation}
J_{s\xi}= \left(  \begin{array}{cc} 1/v_{1\perp} &0\\ 0 &v_{1\perp}
    \end{array} \right)
    \left(  \begin{array}{cc} 1& 0\\ c_1/v_{1\perp} &1
    \end{array} \right)
    \left(  \begin{array}{cc} 1& y_1\\ 0& 1
    \end{array} \right) .
\end{equation}
Its inverse
\begin{equation}
J_{\xi s}\equiv J^{-1}_{s\xi} =
   \left(  \begin{array}{cc} 1& -y_1\\ 0& 1
    \end{array} \right)
    \left(  \begin{array}{cc} 1& 0\\ -c_1/v_{1\perp} &1
    \end{array} \right)
    \left(  \begin{array}{cc} v_{1\perp}& 0\\ 0 &1/v_{1\perp}
    \end{array} \right) ,
\label{J}
\end{equation}
describes the tranformation from $(\delta s_1,\delta v_1)$ to $(\xi ,\kappa )$.

Along similar lines we may derive the Jacobian matrices between $(\delta s_2,
\delta v_2)$ and $(\xi ,\kappa )$. There is a main difference. When following
the above derivation for $s_2$, we use $v_x$ and $v_y$ of the velocity right
before a bounce. By taking this into account, $v_\perp$ in the above formulas 
has to be replaced by $-v_\perp$. For example,
\begin{equation}
\left(\begin{array}{c} \delta s_2 \\ \delta v_2 \end{array} \right)
=\left( \begin{array}{cc}
 -1/v_{2\perp}& -y_2/v_{2\perp}\\ c_2 &-v_{2\perp}+c_2y_2 \end{array} \right)
 \left( \begin{array}{c} \xi \\ \kappa \end{array} \right),
\end{equation}
and 
\begin{equation}
J_{s\xi}(s_2)= \left(  \begin{array}{cc} -1/v_{2\perp} &0\\ 0 &-v_{2\perp}
    \end{array} \right)
    \left(  \begin{array}{cc} 1& 0\\ -c_2/v_{2\perp} &1
    \end{array} \right)
    \left(  \begin{array}{cc} 1& y_2\\ 0& 1
    \end{array} \right) .
\end{equation}
Noting that $l_{12}=y_2-y_1$, we can verify that
$$M(12)=J_{s\xi}(s_2) J_{\xi s}(s_1). $$
For an orbit which starts and ends inside the billiard, respectively, at
$({\bf r}_0,{\bf p}_0)$ and $({\bf r}_t,{\bf p}_t)$, and makes successive
bounces at $s_1$, $s_2$, ..., $s_n$ in between, we have
\begin{equation}
 \left( \begin{array}{c} \xi_t \\ \kappa_t \end{array} \right)
 ={\cal M}\left( \begin{array}{c} \xi_0 \\ \kappa_0 \end{array} \right),
\qquad {\cal M}=J_{\xi s}(s_n)M(n-1,n)\cdots M(12)J_{s\xi}(s_1).
\label{m}
\end{equation}
The Jacobian $\partial \xi_t/\partial \kappa_0$ is determined as
\begin{equation}
\partial \xi_t/\partial \kappa_0 \equiv k \partial r_\perp /\partial
p'_\perp ={\cal M}_{12},
\label{j}
\end{equation}
where $k$ is the absolute value of the conserved momentum.
When all the bounces happen at straight segments of the boundary with
curvature $c=0$, matrix ${\cal M}$ is significantly simplified. In this case,
using expressions (\ref{m}) and (\ref{J}), we have
\begin{equation}
{\cal M}= \left(  \begin{array}{cc} 1& |y_n|\\ 0& 1 \end{array} \right) 
  \left(  \begin{array}{cc} -1& l_{n-1,n}\\ 0& -1 \end{array} \right)\cdots    \left(  \begin{array}{cc} -1& l_{12}\\ 0& -1 \end{array} \right)
  \left(  \begin{array}{cc} 1& |y_1|\\ 0& 1 \end{array} \right)
  =(-1)^n \left(  \begin{array}{cc} 1& L\\ 0& 1 \end{array} \right),
\label{m0}
\end{equation}
where $L$ is the total length of the orbit. Thus, from Eq.~(\ref{j}) we have
\begin{equation}
\partial r_\perp /\partial p'_\perp =(-1)^n L/k
\label{j0}
\end{equation}
for the Jacobian.

{The author thanks Drs.~Y.~Gu, Baowen Li and A.~M.~Ozorio de Almeida for useful 
discussions. This work was supported in part by the National
Natural Science Foundation of China.}



Fig.~1 Images of a given point $Q$ in a corner. Some mediate point $Q'$ is 
used to construct folded paths.

Fig.~2 Perturbation of a path.

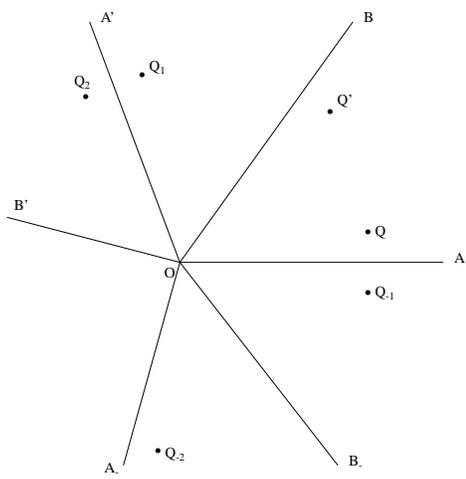
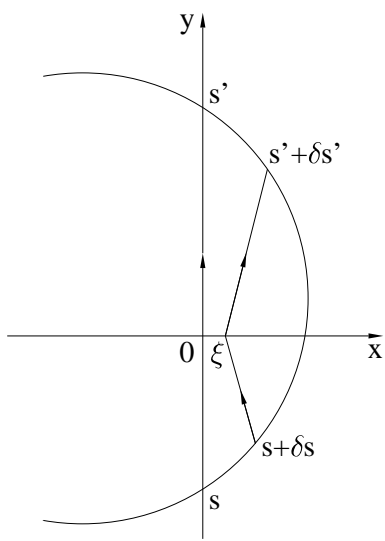
\end{document}